\documentstyle[12pt]{article}
\begin{document}
{\hskip 12.0cm} AS-ITP-96-08\par
\vspace{1.0ex}
\vspace{6ex}
\begin{center}        
{\LARGE \bf QCD Sum Rule Analysis for the 
$\Lambda_b \rightarrow \Lambda_c$ Semileptonic Decay}\\
\vspace{5ex}
{\sc Yuan-ben Dai$^{a}$, Chao-shang Huang$^{a}$, Ming-qiu Huang$^{a}$
and Chun Liu$^{b, a}$}\\
\vspace{3ex}
{\it  $^a$ Institute of Theoretical Physics, Academia Sinica}\\
{\it P.O. Box  2735,  Beijing  100080, China
\footnote{Mailing address}}\\
{\it  $^b$ CCAST (World Laboratory) P.O. Box 8730, Beijing, 100080}\\     
\vspace{8.0ex}
{\large \bf Abstract}\\
\vspace{4ex}
\begin{minipage}{130mm}

   The $1/m_c$ and $1/m_b$ corrections to the 
$\Lambda_b \rightarrow \Lambda_c$ semileptonic decay are analyzed by QCD
sum rules.  Within the framework of heavy quark effective theory, 
the subleading baryonic Isgur-Wise function of 
$\Lambda_b \rightarrow \Lambda_c$ has been calculated.  
It is shown that the corrections due to the $1/m$ Lagrangian insertion are
negligibly small.  The sizable $1/m_Q$ effect to the
decay lies only in the weak current.  The decay spectrum and the 
branching ratio are given.
\par
\vspace{0.5cm}
\end{minipage}
\end{center}

\newpage
      
   The weak decays of heavy baryons provide testing ground for the Standard
Model.  They reveal some important features of the physics of heavy quarks.
From the study of the heavy quark physics, some important parameters of the 
Standard Model, for instance, the Cabbibo-Kobayashi-Maskawa (CKM) matrix 
element $V_{cb}$ can be extracted by comparing experiments with theoretical 
calculations from the decay mode 
$\Lambda_b \rightarrow \Lambda_c l\bar{\nu}$.\par
\vspace{1.0cm}
   The main difficulties in the Standard Model calculations are due to the
poor understanding of the nonperturbative aspects of the strong interactions 
(QCD).  Besides the numerical lattice methods, some analytic, 
model-independent nonperturbative QCD methods have been developed.  For the 
heavy hadrons containing a single heavy quark, an effective theory of QCD 
based on the heavy quark symmetry in the heavy quark limit [1], the 
so-called heavy quark effective theory (HQET), has been proposed [2].  The 
classification of the weak decay form factors of heavy baryons has been
simplified greatly in HQET [3].  To increase the precision of the analysis,
subleading corrections [4] to the results in the heavy quark limit have also 
been considered for baryons [5].  However, for a complete analysis to the 
heavy baryons, we still need to employ some other nonperturbative methods.
\par
\vspace{1.0cm}
   Combining the QCD sum rule [6] method, the complete analysis for heavy 
baryons can be made in HQET.  As a nonperturbative method rooted in QCD 
itself, QCD sum rule has been applied successfully to calculate the 
properties of various hadrons [6, 7].  For the heavy mesons, it has been 
used in the framework of HQET to the leading order heavy quark expansion 
to calculate the masses, the decay constants and the Isgur-Wise function
[8].  And $1/m_Q$ corrections have also been calculated [9, 10].  Heavy
baryons were first calculated by QCD sum rules in Ref. [11].  The heavy
baryon masses and the baryonic Isgur-Wise functions have been calculated 
in the HQET sum rules to the leading order heavy quark expansion in Refs. 
[12] and [13] respectively.  We [14] and another group [15] have calculated 
the $1/m_Q$ corrections to heavy baryon masses of the results of Ref. [12].
In this paper, the subleading Isgur-Wise function of the weak transition
$\Lambda_b \rightarrow \Lambda_c$ is further studied in the HQET sum rules.
\par
\vspace{1.0cm}
    The hadronic matrix element of the weak current 
for $\Lambda_b \rightarrow \Lambda_c$
is parameterized generally by six form factors $F_i$ and $G_i$ 
($i=1, 2, 3$),
\begin{equation}
\begin{array}{lll}
<\Lambda_c(v')|\bar{c}\gamma^{\mu}(1-\gamma^5)b|\Lambda_b(v)>&=&
\bar{u}_{\Lambda_c}(v')(F_1\gamma^{\mu}+F_2v^{\mu}+F_3v'^{\mu})u_{\Lambda_b}
(v)\\
&&-\bar{u}_{\Lambda_c}(v')(G_1\gamma^{\mu}+G_2v^{\mu}+G_3v'^{\mu})\gamma^5
u_{\Lambda_b}(v)~,
\end{array}
\end{equation}
where $v$ and $v'$ denote the four-velocities of $\Lambda_b$ and 
$\Lambda_c$ respectively.  
These form factors need to be determined by some 
nonperturbative QCD method.  
Within the framework of HQET, the classification of them is
simplified very much.  To the order of 
$1/m_Q$, the effective Lagrangian for the heavy quark $h_v$ is
\begin{equation}
\begin{array}{lll}
{\cal L}_{\rm eff}&=&\bar{h}_viv\cdot Dh_v+\frac{1}{2m_Q}{\cal L}'~\\[3mm]
{\cal L}'&=&\bar{h}_v(iD)^2h_v 
-\frac{g}{2}\bar{h}_v\sigma_{\mu\nu}G^{\mu\nu}h_v~.\\[3mm]
\end{array}
\end{equation}
In the heavy quark limit, the form factors are determined 
by only one independent function $\xi(y)$,
\begin{equation}
<\Lambda_c(v')|\bar{h}^{(c)}_{v'}\Gamma h^{(b)}_v|\Lambda_b(v)>=
\xi(y)\bar{u}_{\Lambda_c}(v')\Gamma u_{\Lambda_b}(v)~,
\end{equation}
where $y=v\cdot v'$ and $\Gamma$ is some gamma matrix. 
To the order of $1/m_Q$, they are determined by one mass parameter
$\bar{\Lambda}$ and one additional function $\chi(y)$ which are defined
as follows,
\begin{equation}
\bar{\Lambda}=m_{\Lambda_Q}-m_Q~,
\end{equation}
and
\begin{equation}
<\Lambda_c(v')|{\rm T}\bar{h}^{(c)}_{v'}\Gamma h^{(b)}_v
i\int d^4x\frac{{\cal L}'(x)}{2m_Q}|\Lambda_b(v)>=
\frac{\bar{\Lambda}}{m_Q}\chi(y)
\bar{u}_{\Lambda_c}(v')\Gamma u_{\Lambda_b}(v)~.\\[3mm]
\end{equation}
Both the leading order universal function $\xi$ and the subleading one
$\chi$ are called Isgur-Wise function.
While $\xi$ and  
$\bar{\Lambda}$ have been calculated by the QCD sum rules, we are going 
to calculate the subleading Isgur-Wise function $\chi$.
\par
\vspace{1.0cm}
   QCD sum rule is a calculation method for some nonperturbative physical 
quantities [8].  
The Green's function, from which the Isgur-Wise function can be obtained,
is the three-point 
correlator of the heavy baryonic currents $\tilde{j}$'s and the weak current
in HQET.  Generally the current of the heavy $\Lambda$-baryons is 
\begin{equation}
\tilde{j}^v=\epsilon^{abc}(q_1^{{\rm T}a}C\tilde{\Gamma}\tau q^b_2)h_v^c~,
\end{equation}
where $C$ is the charge conjugate matrix, $\tau$ is an antisymmetric flavor 
matrix, $a, b, c$ denote the color indices, and the choice of the gamma 
matrix $\tilde{\Gamma}$ is not unique, there are two choices,  
\begin{equation}
\tilde{\Gamma}_1=\gamma_5~~~~{\rm and}~~~~
 \tilde{\Gamma}_2=\not v \gamma_5~.
\end{equation}
The current (6) is denoted as $\tilde{j}^v_1$ for $\tilde{\Gamma}_1$ and 
$\tilde{j}^v_2$ for $\tilde{\Gamma}_2$ respectively in the following.
Before performing the sum rule analysis for the three-point correlator, 
which is required to obtain the subleading function $\chi$, let us first 
review some of the two-point correlator results of QCD sum rule [14],
because they are related to the three-point correlator analysis.\par
\vspace{1.0cm}
   In Ref. [14], we obtained the heavy baryon masses and the so-called 
baryonic "decay constants" to the order of $1/m_Q$ by the QCD sum rule 
analysis of some two-point correlators.  With the definition of the 
"decay constant" $f$ in HQET
\begin{equation}
<0|\tilde{j}^v|\Lambda_Q>=f_{\Lambda}u~,\\
\end{equation}
where $u$ is the spinor in HQET, the sum rule gives \footnote{There 
are some errors in the coefficients of the gluon condensates in the 
$1/m_Q$ corrections in Ref. [14].  For $\Lambda$ baryons, the 
coefficients are modified in this paper.  Besides, in Eq. (20) of Ref. 
[14], the coefficients $\frac{13}{3}$ and $\frac{5}{3}$ should be 
replaced by $3$.  However these modifications do not affect the 
numerical results of Ref. [14].}
\begin{equation}
\begin{array}{rcl}
8f^2_{\Lambda_1}e^{-2\bar{\Lambda_1}/T}&=&\displaystyle
\frac{1}{5\times 2^5\pi^4}\int_{0}^{\omega_c}d\omega\omega^5 e^{-\omega/T}
+\frac{4}{3}<\bar{q}q>^2e^{-\frac{m_0^2}{2T^2}}
+\frac{<\alpha_sGG>}{2^4\pi^3}T^2\\[4mm]
&&\displaystyle-\frac{1}{m_Q}(\frac{3}
{5\times 2^7\pi^4}\int_{0}^{\omega_c}d\omega
\omega^6 e^{-\omega/T}+\frac{m_0^2<\bar{q}q>^2}{T}e^{-\frac{m_0^2}{2T^2}}
+\frac{13<\alpha_sGG>}{3\times 2^5\pi^3}T^3)~,\\[4mm]
\end{array}
\end{equation}
for $\tilde{j}^v_1$, and
\begin{equation}
\begin{array}{rcl}
8f^2_{\Lambda_2}e^{-2\bar{\Lambda_2}/T}&=&\displaystyle
\frac{1}{5\times 2^5\pi^4}\int_{0}^{\omega_c}d\omega\omega^5 e^{-\omega/T}
+\frac{4}{3}<\bar{q}q>^2e^{-\frac{m_0^2}{2T^2}}
+\frac{<\alpha_sGG>}{2^4\pi^3}T^2\\[4mm]
&&\displaystyle-\frac{1}{m_Q}(\frac{1}{2^7\pi^4}\int_{0}^{\omega_c}d\omega
\omega^6 e^{-\omega/T}+\frac{m_0^2<\bar{q}q>^2}{T}e^{-\frac{m_0^2}{2T^2}}
+\frac{19<\alpha_sGG>}{3\times 2^5\pi^3}T^3)~,
\end{array}
\end{equation}
for $\tilde{j}^v_2$.  In above equations,  $T$ is the 
Borel parameter.  And 
$\omega_c$ is the continuum threshold.
\par
\vspace{1.0cm} 
   The three-point correlator $\tilde{\Xi}(\omega, \omega', y)$ which we 
choose for sum rule analysis in the HQET is
\begin{equation}
\tilde{\Xi}_{ij}(\omega, \omega', y)=i^2\int d^4x'd^4xe^{ik'x'-ikx}
<0|T\tilde{j}^{v'}_i(x')\bar{h}^{(Q')}_{v'}(0)\Gamma h^{(Q)}_{v}(0)
\bar{\tilde{j}}^v_j(x)|0>~,
~~~~i, j=1,2,
\end{equation}
where $\omega=2v\cdot k$ and $\omega'=2v'\cdot k'$.  
Because of the heavy quark symmetry, $m_Q$ and $m_{Q'}$ are 
taken to be equal for simplicity.  The hadronic representation of this 
correlator is
\begin{equation}
\tilde{\Xi}_{ij}(\omega, \omega', y)=[\frac{4f^2(\xi
+\frac{2\bar{\Lambda}}{m_Q}\chi)}
{(2\bar{\Lambda}-\omega)(2\bar{\Lambda}-\omega')}]_{ij}
\frac{1+\not v'}{2}\Gamma \frac{1+\not v}{2}+{\rm res.}~,
\end{equation}
where $\bar{\Lambda}$ and $f^2$ have been given in the sum rules (9) and
(10) to the order of $1/m_Q$.  On the other hand, 
$\tilde{\Xi}(\omega, \omega', y)$ can be calculated in terms of quark and 
gluon language with vacuum condensates.  This will establish the sum rule.
Only the diagonal correlators ($i=j$) will be 
considered.
It should be remarked here that in general we can consider the correlation
function of the linear combination 
$\tilde{j}^v_1+\tilde{x} \tilde{j}^v_2$ with $\tilde{x}$ being the
mixing parameter.  But with the commonly adopted quark-hadron duality,
the mixed correlator $\tilde{\Xi}_{12}$ has no perturbation
term in the sum rule.  Therefore, the effect due to the mixing is expected 
to be small.
\par
\vspace{1.0cm}
   The calculation of $\tilde{\Xi}(\omega, \omega', y)$ are straightforward.
In addition to the Feynman diagrams at the leading order heavy quark 
expansion which were given in Ref. [13], the diagrams of the $1/m_Q$ 
corrections to the three-point correlator $\tilde{\Xi}(\omega, \omega', y)$ 
are shown in Fig. 1.  They are calculated by including insertions of the 
$1/m_Q$ operators of the Lagrangian (2) with standard method.  The 
chromo-magnetic operator insertion is vanishing for the 
$\Lambda_Q\rightarrow \Lambda_{Q'}$ transition.  Therefore only the kinetic 
energy term insertions need to be considered in our case.  Instead of the 
momentum representation, we adopt the coordinate representation in our 
calculation.  The heavy quark propogator is in a very simple form in the 
coordinate representation so that the calculations become comparatively 
easy.  Taking the insertion of the purely kinetic energy term at the order 
of $1/m_Q$ into account, the heavy quark propogator is 
\begin{equation}
<0|{\rm T}h_v(x)\bar{h}_v(0)|0>=\int_0^{\infty}dt(1-i\frac{t}{2m_Q}
\partial^{\mu}\partial_{\mu})\delta(x-tv)\frac{1+\not v}{2}.
\end{equation}
The fixed point gauge [16] is used.  
All the condensates with dimensions 
lower than 6 are retained.  We also include the dimension 6 condensate 
$<\bar{q}(x)q(x')>^2$ in our analysis which is a main contribution.  We 
use the gaussian ansatz for the distribution in spacetime for this 
condensate [17].  We use the following values of the condensates,
\begin{equation}
\begin{array}{rcl}
<\bar{q}q>&\simeq&-(0.23~ {\rm GeV})^3~,\\
<\alpha_sGG>&\simeq&0.04~ {\rm GeV}^4~,\\
<g\bar{q}\sigma_{\mu\nu}G^{\mu\nu}q>&\equiv&m_0^2<\bar{q}q>~,
~~~~~~m_0^2\simeq0.8~{\rm GeV}^2~.\\
\end{array}
\end{equation} 
The normalization ${\rm Tr} \tau^{\dagger}\tau=1$ has been used in the 
analysis.  
In the fixed-point gauge, the space-time 
translational invariance is violated, but it is restored by adding all 
the diagrams in Fig. 1.  This is a check of our calculation.\par
\vspace{1.0cm}
   We use the commonly adopted quark-hadron duality for the resonance 
part of Eq. (12).  Generally 
the duality is to simulate the resonance contribution by the perturbative 
part above some threshold energy $\omega_c$.  The perturbative 
contribution of the three-point correlator 
$\tilde{\Xi}^{\rm pert}(\omega, \omega', y)$ can be expressed by the 
dispersion relation, 
\begin{equation}
\tilde{\Xi}^{\rm pert}(\omega, \omega', y)=\frac{1}{\pi}\int_0^{\infty}
d\tilde{\omega}\int_0^{\infty}d\tilde{\omega}'
\frac{{\rm Im}\tilde{\Xi}^{\rm pert}(\tilde{\omega}, \tilde{\omega}', y)}
{(\tilde{\omega}-\omega)(\tilde{\omega'}-\omega')}.
\end{equation}
The integration domain is a kitelike area.  With the redefinition of the 
integral variables,
\begin{equation}
\begin{array}{lll}
\tilde{\omega}_+&=&\displaystyle
\frac{\tilde{\omega}+\tilde{\omega}'}{2}~,\\[3mm]
\tilde{\omega}_-&=&\displaystyle
(\frac{y+1}{y-1})^{1/2}
\frac{\tilde{\omega}-\tilde{\omega}'}{2}~,\\[3mm]
\end{array}
\end{equation}
the integration becomes
\begin{equation}
\int_0^{\infty}d\tilde{\omega}\int_0^{\infty}d\tilde{\omega}' \ldots
=2(\frac{y-1}{y+1})^{1/2}
\int_0^{\infty}d\tilde{\omega_+}
\int_{-\omega_+}^{\omega_+}d\tilde{\omega}_- \ldots~.\\[3mm]
\end{equation}
It is in $\omega_+$, that the quark-hadron duality is assumed [18],
\begin{equation}
{\rm res.}=\frac{2}{\pi}(\frac{y-1}{y+1})^{1/2}
\int_{\omega_c}^{\infty}d\tilde{\omega_+}
\int_{-\omega_+}^{\omega_+}d\tilde{\omega}_-
\frac{{\rm Im}\tilde{\Xi}^{\rm pert}(\tilde{\omega}, \tilde{\omega}', y)}
{(\tilde{\omega}-\omega)(\tilde{\omega'}-\omega')}~.\\[3mm]
\end{equation}
\par
\vspace{1.0cm} 
   In the heavy quark limit, we have double checked the analysis of Ref. 
[13].  There are two sum rules for the leading order Isgur-Wise function 
corresponding to two choices of the baryonic current.  When $\omega_c$ 
lies between $1.8-2.5$ GeV, the stability window of $T$ exists,
$T=0.3-0.6$ GeV.  The two results for the Isgur-Wise function are 
consistent with each other.  
For $y$ lies in the physical region $1 - 1.43$, the 
linear approximation can fit the results,
\begin{equation}
\xi(y)=1-\rho(y-1)~,~~~~~\rho=0.55\pm0.15~,
\end{equation}
where the uncertainty of $\rho$ accounts those of $\omega_c$ and $T$, in
addition to the difference of the two sum rule results.  For $y$ lies in 
$1-3$, we find that the following function fit very well to
our numerical results for  
the Isgur-Wise function for reasonable $\omega_c$ and $T$,
\begin{equation}
\xi(y)=(\frac{2}{y+1})^{0.5}\exp(-0.8~\frac{y-1}{y+1})~.\\[3mm]
\end{equation}
We note that the $y$-dependence of the Isgur-Wise function is not as 
steep as that of the Skyrme model [19] and the quark 
model [20].\par
\vspace{1.0cm}
   The sum rule for the subleading Isgur-Wise function $\chi(y)$ is
\begin{equation}
\chi(y)=-\frac{e^{2\bar{\Lambda}/T}}
{8\bar{\Lambda}f^2}[J(y)-\xi(y)J(1)]~,
\end{equation}
where
\begin{equation}
\begin{array}{lll}
J_1(y)&=&\displaystyle(\frac{1}{2\pi}\frac{1}{y+1})^4\frac{4y-1}{5}
\int_{0}^{\omega_c}d\omega \omega^6 e^{-\omega/T}
+\frac{m_0^2<\bar{q}q>^2}{6T}\\[4mm]
&&\displaystyle \cdot [3+\frac{m^2_0}{4T^2}(y^2-1)]
e^{-\frac{m_0^2}{4T^2}(y+1)}+\frac{<\alpha_sGG>}{3}(\frac{1}{2\pi}
\frac{T}{y+1})^3(4y^2+3y+6)~,\\[4mm]
J_2(y)&=&\displaystyle(\frac{1}{2\pi}\frac{1}{y+1})^4y
\int_{0}^{\omega_c}d\omega \omega^6 e^{-\omega/T}
+\frac{m_0^2<\bar{q}q>^2}{6T}y[3+\frac{m^2_0}{4T^2}(y^2-1)]
e^{-\frac{m_0^2}{4T^2}(y+1)}\\[4mm]
&&\displaystyle +\frac{<\alpha_sGG>}{3}(\frac{1}{2\pi}
\frac{T}{y+1})^3(2y^3+8y^2+4y+5)~,\\[4mm]
\end{array}
\end{equation}
with the subscripts 1 and 2 denoting the two kinds of baryonic currents.
The Luke's theorem [4] in the baryon case $\chi(1)=0$ is satisfied 
automatically.  The numerical results are shown in Fig. 2 where the two
curves correspond to the two sum rule results.  The range of $\omega_c$ 
is the same as that in the leading order.  The sum rule window is 
narrower than the leading order one.  In the window $T=0.35-0.55$ the 
results for the subleading Isgur-Wise function are stable.  The two 
results can also be regarded as being consistent with each other.  
Nevertheless, it is obvious that the subleading Isgur-Wise function is 
negligibly small,
\begin{equation}
\chi(y)\simeq O(10^{-2})~.
\end{equation}
\par
\vspace{1.0cm}
   The semileptonic decay $\Lambda_b\rightarrow\Lambda_cl\bar{\nu}$ can 
be analyzed directly after obtaining the hadronic matrix elements from 
the QCD sum rules.  
By neglecting the lepton mass, it is easy to show
that the differential decay rate is
\begin{equation}
\begin{array}{lll}
\displaystyle\frac{1}{\sqrt{y^2-1}}
\frac{d\Gamma(\Lambda_b \rightarrow \Lambda_c l \bar{\nu})}{dy}&=&
\displaystyle
\frac{G_F^2|V_{cb}|^2m_{\Lambda_b}^2m_{\Lambda_c}^3}{(2\pi)^3}
\{(1-2ry+r^2)[(y-1)F_1^2+(y+1)G_1^2]\\
&&\displaystyle +\frac{y^2-1}{3}(Ar^2+2Br+C)\}~,
\end{array}
\end{equation}
where $r=m_{\Lambda_c}/m_{\Lambda_b}$.  In the above equation,
\begin{equation}
\begin{array}{lll}
A&=&2F_1F_2+(y+1)F_2^2+2G_1G_2+(y-1)G_2^2~,\\
B&=&F_1^2+F_1F_2+F_2F_3+F_3F_1+yF_2F_3+G_1^2-G_1G_2-G_2G_3+G_3G_1+yG_2G_3~,\\
C&=&(y+1)F_3^2+2F_1F_3+(y-1)G_3^2-2G_1G_3~.\\[3mm]
\end{array}
\end{equation}
To the order of both $1/m_c$ and $1/m_b$, the form factors $F_i$ and 
$G_i$ are expressed as
\begin{equation}
\begin{array}{lll}
F_1&=&\displaystyle C(\mu)\xi(y)+(\frac{\bar{\Lambda}}{2m_c}
+\frac{\bar{\Lambda}}{2m_b})[2\chi(y)+\xi(y)]~,\\[3mm]
G_1&=&\displaystyle C(\mu)\xi(y)+(\frac{\bar{\Lambda}}{2m_c}
+\frac{\bar{\Lambda}}{2m_b})[2\chi(y)+\frac{y-1}{y+1}\xi(y)]~,\\[3mm]
F_2&=&\displaystyle G_2=-\frac{\bar{\Lambda}}{m_c(y+1)}\xi(y)~,\\[3mm]
F_3&=&\displaystyle -G_3=-\frac{\bar{\Lambda}}{m_b(y+1)}\xi(y)~,\\[3mm]
\end{array}
\end{equation}
where $C(\mu)$ is the perturbative QCD coefficient.
The subleading 
Isgur-Wise function can be safely neglected.  The $1/m_Q$ corrections 
are mainly due to the weak current.  With the form of the leading order
Isgur-Wise function (19), the differential decay rate of 
$\Lambda_b\rightarrow\Lambda_cl\bar{\nu}$ is shown in Fig. 3.  In 
Fig. 3, we have taken the heavy quark masses $m_b=4.83$ GeV, 
$m_c=1.44$ GeV
and $\bar{\Lambda}=0.79$ GeV [14], the renormalization point 
$\mu=470$ MeV, the CKM matrix element $V_{cb}=0.04$ [21].  The
width and the  
branching ratio of this decay mode are
\begin{equation}
\begin{array}{lll}
{\rm \Gamma}&=&6.05\times10^{-14}~{\rm GeV}~,\\
{\rm Br}&=&9.8\%~.
\end{array}
\end{equation}
The $1/m_Q$ correction possesses $10\%$ in the above branching ratio.
\par
\vspace{1.0cm}
   We have analyzed the $\Lambda_b\rightarrow\Lambda_c$ semileptonic 
decays by QCD sum rules within the framework of HQET to the order of 
$1/m_c$ and $1/m_b$.  In the heavy quark limit, the analysis for the
$\Lambda_b\rightarrow\Lambda_c$ decay depends on one independent form 
factor which is the leading order Isgur-Wise function and was calculated 
in the QCD sum rules [13].  However, for a more precise analysis, only
leading order calculation is not enough.  In this paper, we have 
considered the $1/m_Q$ corrections.  The subleading Isgur-Wise function
has been calculated by the HQET sum rules.  It is shown to be so small 
that it can be neglected.  
The $1/m_Q$ correction to the decay $\Lambda_b\rightarrow\Lambda_c$
results only from the weak current.  The decay differential distribution
has been given.  The branching ratio is predicted to be 
${\rm Br}(\Lambda_b\rightarrow\Lambda_cl\bar{\nu})=9.8\%$  after taking
$V_{cb}=0.04$.  This will be useful to the experiments in the near 
future.  The polarization effects of this decay have not been calculated
which will be considered elsewhere.\par
\vspace{1.0cm}  
   Finally we would like to make a remark on the perturbative QCD 
corrections in the sum rule calculations.  Such corrections to the 
baryonic Isgur-Wise function which still have not been included, would 
involve us in the three-loop calculations.  However, we expect that they 
should be small.  The Isgur-Wise function obtained from the QCD sum rule 
actually is a ratio of the three-point correlator to the two-point 
correlator results.  While both of these correlators subject to large 
perturbative QCD corrections, their ratio does not depend on these 
corrections significantly because of cancelation.  Therefore the results 
for the Isgur-Wise function are more reliable than that for the heavy 
baryon masses.  This is what happened in the heavy meson case [8].  The 
perturbative QCD corrections to the two-point correlators, therefore to 
the heavy baryon masses, will be calculated elsewhere.\par

\vspace{2.0cm}
\begin{center}
{\large \bf Acknowledgement}\par
\end{center}
\vspace{1.0cm}
   This work is supported in part by the National Science Foundation
of China.  Liu is supported by the China Postdoctoral Science 
Foundation.

\newpage

{\Large \bf Figure captions}\\

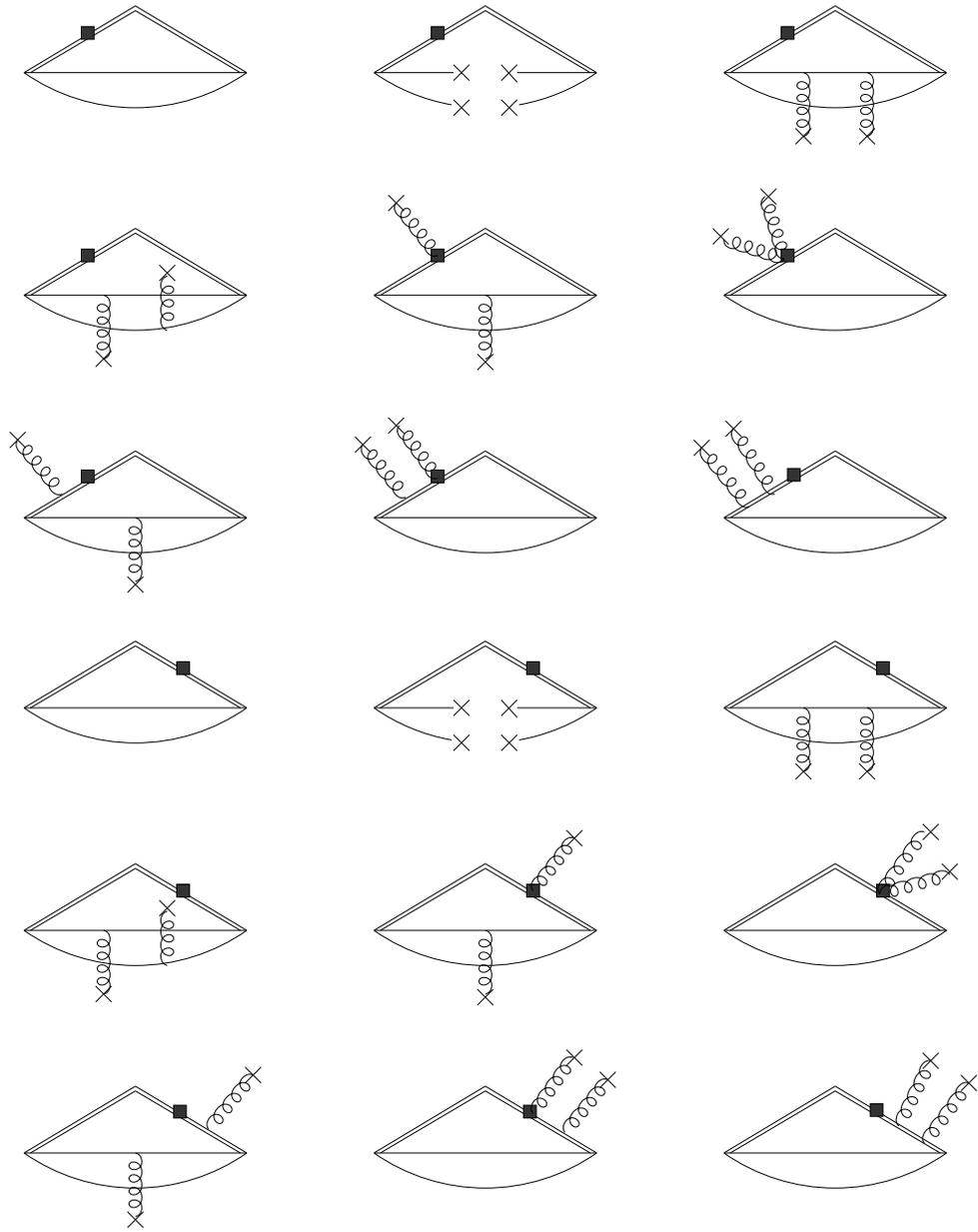
Fig. 1.  Feynman diagrams for the $1/m_Q$ corrections to 
$\tilde{\Xi}(\omega, \omega', y)$.  The insertions are only the kinetic 
energy terms at the order of $1/m_Q$.\\

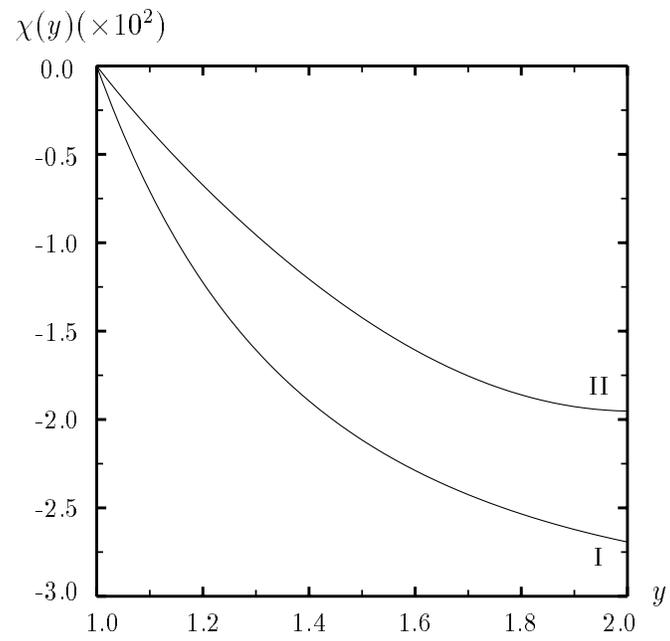
Fig. 2.  Subleading Isgur-Wise function $\chi(y)$.  The lower and the 
upper curves correspond to the sum rules (29) of $J=J_1$ 
with $\omega_c=2.2$ GeV, $T=0.55$ GeV
and $J_2$ with $\omega_c=2.5$ GeV,  $T=0.39$ GeV
respectively.\\

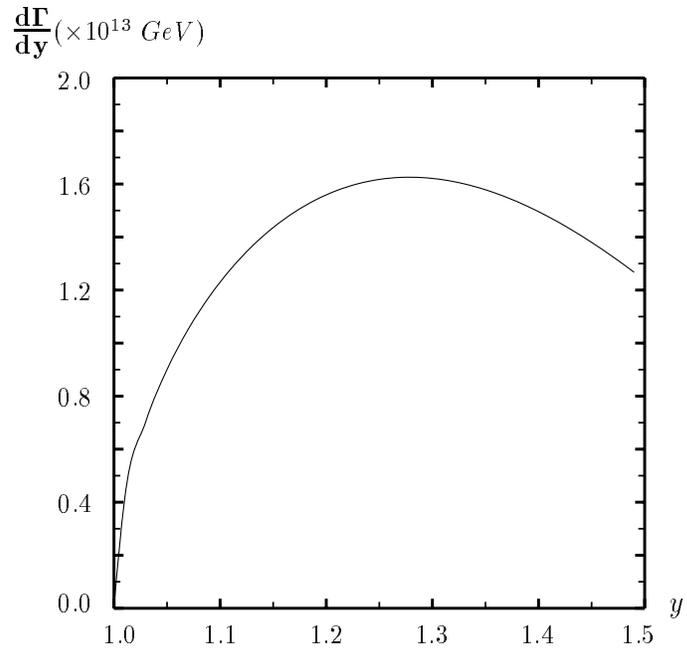
Fig. 3.  The differential decay rate of 
$\Lambda_b\rightarrow \Lambda_c l\bar{\nu}$. ($y=v\cdot v'$)

\end{document}